\documentclass[lettersize,journal]{IEEEtran}
\usepackage{amsmath,amsfonts}
\usepackage{algorithmic}
\usepackage{algorithm}
\usepackage{array}
\usepackage[caption=false,font=normalsize,labelfont=sf,textfont=sf]{subfig}
\usepackage{textcomp}
\usepackage{stfloats}
\usepackage{url}
\usepackage{multirow}
\usepackage{booktabs}
\usepackage{verbatim}
\usepackage{graphicx}
\usepackage{cite}
\usepackage{caption2}
\usepackage{subfiles}
\usepackage{makecell}
\begin{document}

\title{A Low-Cost, High-Precision Human-Machine Interaction Solution Based on Multi-Coil Wireless Charging Pads}

\author{
    Bojun Zhang\IEEEauthorrefmark{1} 
    \thanks{
    \IEEEauthorrefmark{1}Bojun Zhang is the corresponding author. Tianjin University, China Tianjin 300000, China. 
    Email: ewwllvraier@126.com
}
}

\maketitle
\thispagestyle{plain}
\begin{abstract}
Wireless charging pads are common, yet their functionality is mainly restricted to charging. Existing gesture recognition techniques, such as those based on machine vision and WiFi, have drawbacks like high costs and poor precision. This paper presents a new human - machine interaction solution using multi - coil wireless charging pads.
The proposed approach leverages the pads' existing modules without additional wearable sensors. It determines gestures by monitoring current and power changes in different coils. The data processing includes noise removal, sorting, high - pass filtering, and slicing. A Bayesian network and particle filtering are employed for motion tracking. 
Through experiments, this solution proves to have wide applications, high recognition accuracy, and low cost. It can effectively identify diverse gestures, increasing the value of wireless charging pads. It outperforms traditional methods, with a 0.73 improvement in recognition accuracy and better environmental adaptability.
\end{abstract}


\section{Introduction}
\subsection{Research Background}
With the rapid advancement of technology, wireless charging has become an indispensable part of modern life. Currently, wireless charging pads are widely used, and billions of dollars' worth of wireless charging pads have been deployed in homes and public places. However, despite the high penetration rate of wireless charging pads, their functionality is primarily limited to charging purposes, which restricts the potential and scope of wireless charging pads and fails to fully leverage their convenience and multifunctionality in daily life. This study aims to enhance the utilization rate of wireless charging pads and bring more convenience to users by adding interactive functions to existing wireless charging pads at a low cost. By achieving this goal, we can not only improve the efficiency of wireless charging pads but also expand their application prospects in smart homes, office automation, and public spaces.

The expansion of interactive functions of wireless charging pads will bring about the following potential application scenarios: In home environments, users can turn on all the lights in the house by making a gesture over the wireless charging pad, achieving convenient control of smart homes; In public cafes and other venues, users can adjust the lighting above their heads by swiping on the wireless charging pad, enhancing the user experience in public spaces; In office environments, employees can enter work mode by making a gesture over the wireless charging pad, simplifying work processes and improving work efficiency. These application scenarios demonstrate the potential of wireless charging pads in human-machine interaction and their significant role in enhancing the convenience of daily life. Through this research, we hope to provide new ideas and solutions for the multifunctional application of wireless charging pads and the development of human-machine interaction technology.
\subsection{Related work}
Gesture recognition algorithms can be categorized into machine vision based algorithms, wireless signal based algorithms and wearable device based algorithms depending on the input signal\cite{wang2018,oudah2020,liu2024wireless}.
\subsubsection{Machine Vision based Algorithms}
Wan and Van Gool\cite{wan2016eccv} proposed a method for hand pose estimation based on local surface normals. This method analyzes the local surface normal information of objects to effectively estimate the hand pose.
Li et al.\cite{li2016iccv} achieved 3D hand pose estimation using a randomized decision forest with segmentation index points. By constructing a randomized decision forest model and combining specific segmentation index points, the accuracy of 3D hand pose estimation is effectively improved.
Algorithms based on computer vision, while providing rich visual information, suffer from high deployment costs, privacy concerns, and large video data processing volumes that affect system real-time performance.
\subsubsection{Wireless based Algorithms}
Li et al.\cite{li2015mobicom} introduced the WiDraw technology, aiming to achieve hands - free drawing in the air using commercial WiFi devices. This technology analyzes WiFi signals to recognize users' hand gestures in the air, enabling contact - free drawing operations and expanding the application of WiFi technology in the field of human - machine interaction.
Wang et al.\cite{he2015wig} proposed a gesture recognition control system based on WiFi signals. It processes the signals according to the changes in user gestures to obtain specific information about the gestures, and then matches this information with pre - set commands to achieve the control of smart homes and household appliances. 
Algorithms based on wireless signals, such as WiFi positioning, have low precision and limited application scenarios, especially when the user's hand blocks signals from a certain direction, causing the signal strength of the corresponding angle to decrease. 
\subsubsection{Wearable Device based Algorithms}
Shen et al.\cite{shen2021measurement} proposed a method for full - pose estimation using inertial and magnetic sensor fusion in a structured magnetic field for hand motion tracking. This method combines multiple sensor data to improve the accuracy and comprehensiveness of hand pose tracking.
Alexander and Perry\cite{alexander2001spie} studied the technology of magnetic localization and real - time tracking of concealed threats. By utilizing the characteristics of the magnetic field, it realizes the localization and real - time tracking of potential threat targets, which has important application value in the field of security monitoring.
Wearable device-based algorithms, which use IMU to judge the movement of joints based on changes in magnetic force in the magnetic field, can provide relatively accurate gesture recognition but still require additional sensors and magnetic fields, limiting their application scope.

In contrast, the gesture recognition interaction system based on wireless charging pads proposed in this study has clear application advantages. This system requires no additional devices, has high recognition accuracy, and is cost-effective, suitable for a wide range of application scenarios. This system not only improves the utilization rate of wireless charging pads but also provides new interaction methods for smart homes, office automation, and public spaces, with great potential for development and application prospects.
\subsection{Challenges and Solutions}
In the context of our specific application scenarios, we identified two key challenges that our proposed human-machine interaction solution needed to address. Correspondingly, we employed Bayesian networks and particle filtering as our solutions to tackle these challenges.

\textbf{Challenge in Dynamic Gesture Recognition.}
The first challenge arises from the dynamic nature of gesture recognition in various environments. The variability in user movements and the complexity of gesture dynamics pose significant difficulties for accurate recognition. To address this, we utilized particle filtering to model the uncertainty and non-linearity in gesture trajectories. Particle filtering allows us to infer the motion trajectory of user gestures effectively, providing a robust method for handling the dynamic changes in the system state.

\textbf{Challenge in Complex Data Relationships.}
The second challenge is related to the complex relationships within the data collected from the multi-coil wireless charging pads. The data, influenced by multiple factors including user gestures and environmental interactions, requires sophisticated modeling to extract meaningful insights. We addressed this challenge by implementing Bayesian networks, which are adept at capturing probabilistic dependencies and modeling complex relationships within the data. This approach enables us to accurately interpret the electromagnetic variations induced by user interactions, enhancing the gesture recognition process.

By applying these solutions, we have effectively mitigated the challenges inherent to our application scenarios. The use of particle filtering for dynamic gesture recognition and Bayesian networks for modeling complex data relationships has significantly improved the performance of our human-machine interaction solution.
\section{Overview}
\subsection{System Overview}
\begin{figure}[htbp]
    \centering
    \includegraphics[width=0.45\textwidth]{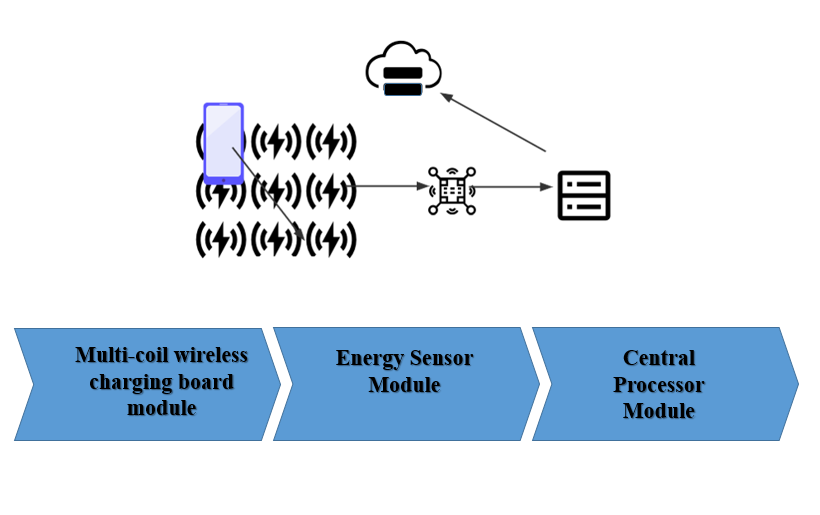} 
    \caption{System Workflow Diagram}
    \label{fig:system_overview}
\end{figure}
This system aims to enhance the functionality of existing multi-coil wireless charging pads with human-machine interaction capabilities at a low cost. The system mainly consists of the following components:

\begin{itemize}
    \item \textbf{Energy Sensor Module}: Responsible for detecting real-time current and voltage changes in the magnetic induction coils and transmitting the data to the central processing unit in real-time.
    \item \textbf{Multi-Coil Wireless Charging Module}: Composed of coils and a controller, it activates and causes current changes when a smart device passes overhead.
    \item \textbf{Central Processing Unit}: Receives data from the energy sensor module and uses machine learning algorithms to process and analyze the data, recognizing user gestures.
\end{itemize}

The workflow of the system is shown in Figure \ref{fig:system_overview}. The energy sensor module first captures current and voltage data from the coils, which is then transmitted to the central processing unit. The central processing unit identifies user gestures by analyzing the current and power changes in different coils and uses the recognition results to control corresponding operations of smart devices.

With this design, the system not only improves the utilization rate of wireless charging pads but also provides a new, low-cost, and high-precision solution for human-machine interaction technology. The system has broad application prospects in areas such as smart homes, office automation, and public spaces\cite{liu2024fine}.
\begin{figure*}[htbp]
    \centering
    \includegraphics[width=0.9\textwidth]{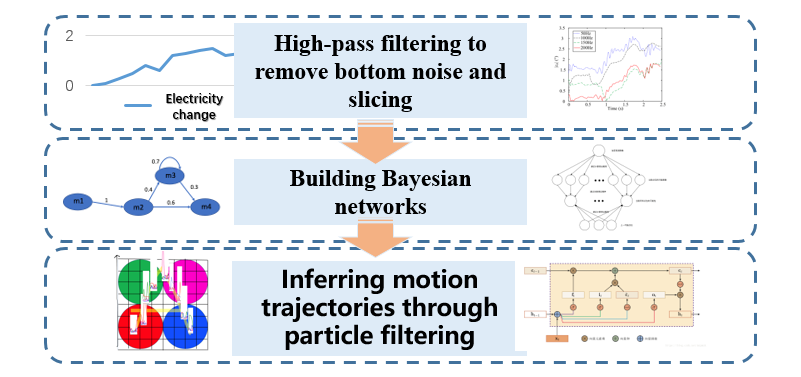} 
    \caption{Workflow of Data Processing and Gesture Recognition in the Central Processing Unit}
    \label{fig:data_processing_workflow}
\end{figure*}
\subsection{Energy Sensor Module}
The Energy Sensor Module is a critical component of our system, responsible for capturing real-time current and voltage fluctuations from the magnetic induction coils. This module plays an essential role in the gesture recognition process, as the variations in current and voltage are directly correlated with the user's hand movements.
\subsubsection{Module Description}
\begin{figure}[htbp]
    \centering
    \includegraphics[width=0.35\textwidth]{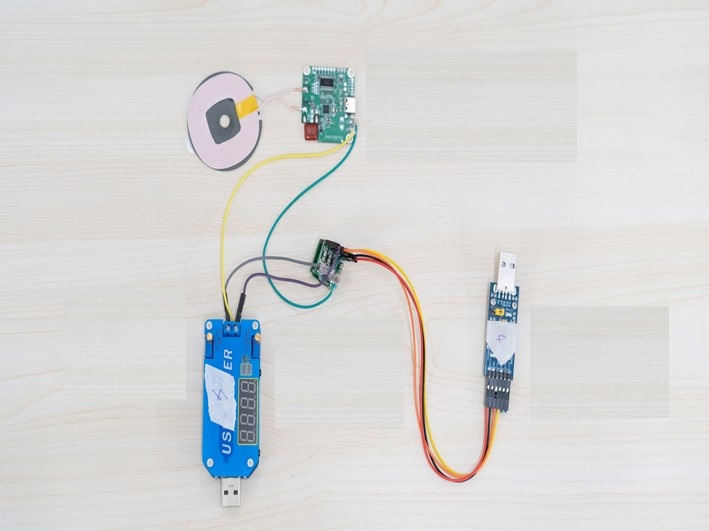} 
    \caption{Energy Sensor Module Interfaced with Multi-Coil Wireless Charging Pad}
    \label{fig:energy_sensor_module}
\end{figure}
The Energy Sensor Module is designed to interface with multiple coils simultaneously, ensuring that data from all coils is collected without interference. It consists of several key elements:
\begin{itemize}
    \item Current and voltage sensors to measure the electrical parameters from each coil.
    \item A microcontroller unit (MCU) for processing the sensor data and managing communication with the Central Processing Unit.
    \item An amplifier circuit to boost the sensor signals for more accurate readings.
    \item A filtering circuit to remove noise and ensure data integrity.
\end{itemize}
\subsubsection{Operation}
The module operates by continuously monitoring the current and voltage across the coils. When a user's hand interacts with the wireless charging pad, it causes changes in the electromagnetic field, which in turn affects the current flowing through the coils. These changes are detected by the sensors and processed by the MCU. The processed data is then sent to the Central Processing Unit for further analysis and gesture recognition.
\subsubsection{Integration with the System}
The Energy Sensor Module is integrated into the system such that it is connected to each coil via dedicated wiring. This setup ensures that every coil's data is independently captured and transmitted to the Central Processing Unit. The module is also designed to work in conjunction with the Multi-Coil Wireless Charging Module, leveraging the existing infrastructure of the wireless charging pad to enhance its functionality.

Figure \ref{fig:energy_sensor_module} illustrates the Energy Sensor Module interfaced with the Multi-Coil Wireless Charging Pad. The image shows the physical setup where the sensor modules are connected to the coils. Each sensor module is strategically placed to correspond with a coil, ensuring optimal data capture for gesture recognition.

This detailed setup of the Energy Sensor Module not only enhances the system's ability to accurately interpret user gestures but also demonstrates the practical implementation of the system's design principles.
\subsection{Multi-Coil Wireless Charging Module}
The Multi-Coil Wireless Charging Module is the backbone of our human-machine interaction system, providing the necessary infrastructure for both wireless power transfer and gesture recognition. This module is composed of multiple coils and a controller, which work in unison to detect the presence of a smart device and initiate the charging process, as well as to capture the electromagnetic changes that correspond to user gestures.
\subsubsection{Module Composition}
The module consists of:
\begin{itemize}
    \item Multiple wireless charging coils arranged in a specific pattern to cover a wide area and ensure efficient energy transfer.
    \item A controller unit that manages the charging process and coordinates with the Energy Sensor Module to capture gesture data.
    \item Connectivity interfaces that link the coils and controller to the Energy Sensor Module and the Central Processing Unit.
\end{itemize}
\begin{figure}[htbp]
    \centering
    \includegraphics[width=0.35\textwidth]{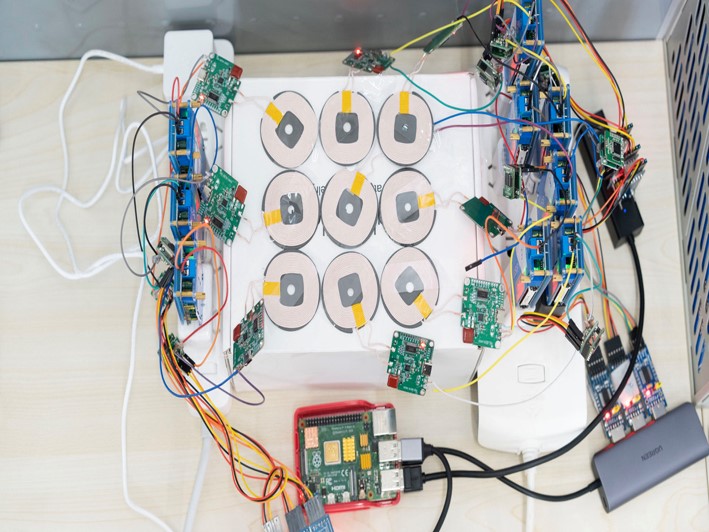} 
    \caption{Multi-Coil Wireless Charging Module with Integrated Sensors}
    \label{fig:multi_coil_module}
\end{figure}
\subsubsection{Functionality}
The functionality of the Multi-Coil Wireless Charging Module extends beyond mere charging. When a smart device is placed on or passes over the charging pad, the coils induce a current in the device's receiving coil, initiating the charging process. Simultaneously, any movement or gesture by the user affects the electromagnetic field, causing variations in the coil's current. These variations are detected by the sensors and sent to the Central Processing Unit for analysis.

\subsubsection{Integration and Data Flow}
The module is integrated into the system such that each coil is connected to both the controller and the Energy Sensor Module. The data flow involves the coils detecting changes in the electromagnetic field, which are then processed by the controller and transmitted to the Energy Sensor Module. The processed data is further analyzed in the Central Processing Unit to interpret user gestures.

This module's dual functionality of charging and gesture detection makes it a versatile component for smart environments, enhancing the wireless charging pad's capabilities without requiring additional hardware.
\subsection{Central Processing Unit}
The Central Processing Unit (CPU) serves as the analytical engine of our system, tasked with the crucial role of interpreting complex data from the Energy Sensor Module to identify and classify user gestures. This unit implements a sophisticated data processing strategy that begins with high-pass filtering to remove low-frequency noise and slice the data into manageable segments. Following this initial cleaning, a Bayesian network is constructed to model the probabilistic relationships within the data, and particle filtering is applied to infer motion trajectories. As depicted in Figure \ref{fig:data_processing_workflow}, this integrated approach allows for accurate gesture recognition by analyzing the electromagnetic changes induced by user interactions with the multi-coil wireless charging pad.

This overview highlights the CPU's pivotal role in transforming raw sensor data into interpretable gestures, showcasing the system's capability for precise and efficient human-machine interaction.
\subsubsection{Data Processing}
The initial phase of data processing involves several crucial steps aimed at preparing the raw data for subsequent analysis and gesture recognition. This section details the methodologies employed for noise reduction, filtering, and segmentation of the data.

The raw data collected from the coils contains various sources of noise that can obscure meaningful patterns. To mitigate this, the data undergoes an initial cleaning process:
\begin{itemize}
    \item \textbf{Noise Reduction}: Techniques such as moving average or median filtering are applied to smooth out high-frequency noise components.
    \item \textbf{Temporal Sorting}: The data points are sorted in chronological order to ensure that the temporal sequence of events is preserved.
\end{itemize}
High-pass filtering is a critical step in the data processing pipeline, designed to remove low-frequency components, often referred to as "bottom noise". This step is essential for enhancing the signal-to-noise ratio and focusing on the relevant frequency bands that carry information about user gestures. Mathematically, the high-pass filter can be represented as:
\begin{equation}
    y[n] = b_1 x[n] + b_2 x[n-1] - a_1 y[n-1]
\end{equation}
where \(x[n]\) is the input signal, \(y[n]\) is the filtered output, and \(b_1, b_2, a_1\) are the filter coefficients determined by the desired cutoff frequency.

Following the high-pass filtering, the data is segmented into manageable chunks to facilitate analysis. Each segment corresponds to a set of five consecutive acquisitions, allowing for the examination of gesture dynamics over discrete time intervals. This segmentation is crucial for applying machine learning models that operate on fixed-size input vectors.

\begin{figure}[htbp]
    \centering
    \includegraphics[width=0.42\textwidth]{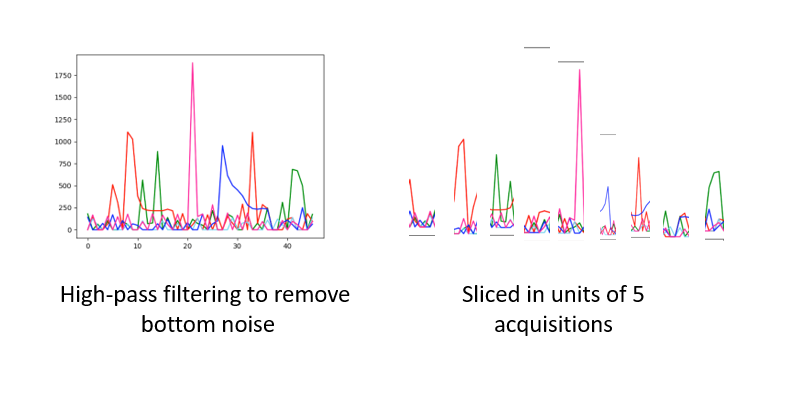} 
    \caption{High-pass filtering and data segmentation}
    \label{fig:high_pass_filtering}
\end{figure}

Figure \ref{fig:high_pass_filtering} illustrates the effect of high-pass filtering and subsequent segmentation. The left graph shows the raw data with significant noise, while the right graph displays the data after filtering and segmentation, highlighting the clarity and structure brought about by these preprocessing steps.

These initial processing steps are foundational for the subsequent stages of gesture recognition, ensuring that the data fed into the models is of high quality and relevance.
\subsubsection{Model Construction}
This section details the construction of our model for localization and trajectory inference using multi - coil wireless charging pads. Our approach leverages particle filtering\cite{arulampalam2002} and Bayesian networks\cite{pearl1988} to achieve precise gesture recognition and tracking.

Particle filtering is a recursive Bayesian estimation algorithm that is particularly useful for systems with non - linear dynamics and/or measurements. It represents the probability distribution of the system state by a set of random samples, known as particles, and their associated weights.
The prediction step is given by:
\begin{equation}
    \hat{x}_{t|t - 1} \sim f(\hat{x}_{t - 1|t - 1}, u_t)
\end{equation}
where \(\hat{x}_{t|t - 1}\) is the predicted state and \(u_t\) is the control input.

The weight calculation in the update step is:
\begin{equation}
    w_t \propto p(z_t | \hat{x}_{t|t - 1})
\end{equation}
where \(z_t\) is the measurement.

The weight normalization in the update step is:
\begin{equation}
    w_t \leftarrow \frac{w_t}{\sum_{i = 1}^N w_i}
\end{equation}

The resampling step is:
\begin{equation}
    \hat{x}_{t|t} \sim \sum_{i = 1}^N w_i \delta(\hat{x}_{t|t} - \hat{x}_{i,t|t - 1})
\end{equation}
where \(\delta\) is the Dirac delta function.

A Bayesian network is employed to model the probabilistic relationships between the system's variables. It is particularly useful for capturing complex dependencies and for making predictions based on observed data.

Let \(P\) denote the probability, \(x_{t}\) represent the current position (\(\text{now\_position}\)), \(x_{t - 1}\) represent the last position (\(\text{last\_position}\)), and \(\lambda\) represent the eigenvalue (\(\text{eigenvalue}\)). The Bayesian estimator is denoted as \(BE\), and the variable elimination operation is denoted as \(VE\).
The probability of the current position given the last position and eigenvalue is calculated by:
\begin{equation}
    P(x_{t} | x_{t - 1}, \lambda) = BE(x_{t - 1}, \lambda)
\end{equation}

The posterior probability is computed by:
\begin{equation}
    P_{\text{posterior}} = VE(x_{t - 1}, x_{t}, \lambda)
\end{equation}

The posterior probability distribution obtained from the Bayesian network serves as the state transition matrix in the particle filter, providing a robust method for handling uncertainties and non - linearities in the system.

\begin{figure}[htbp]
    \centering
    \includegraphics[width = 0.42\textwidth]{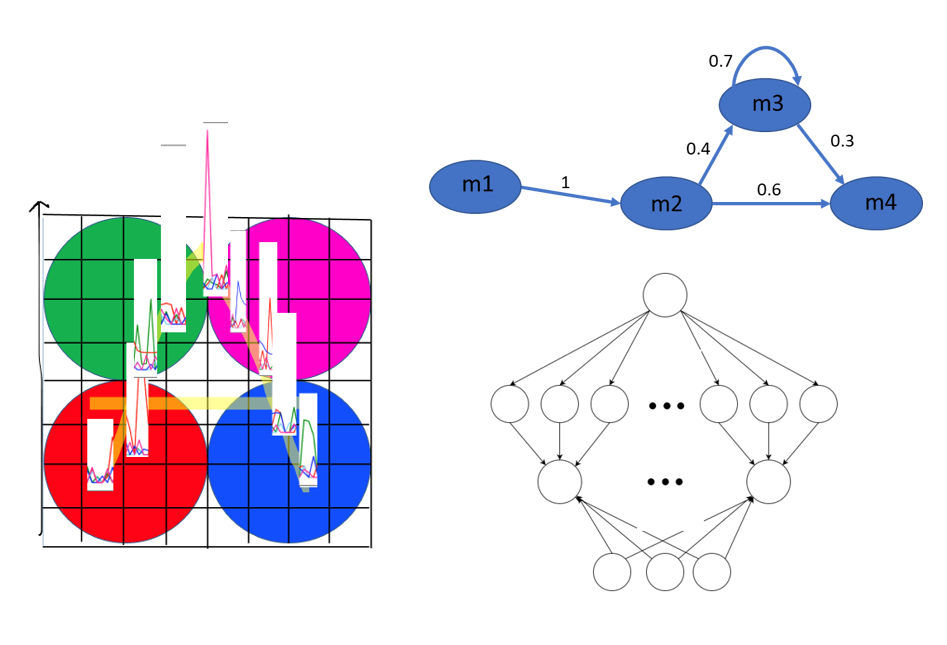} 
    \caption{Visualization of the Bayesian Network and Particle Filtering Process}
    \label{fig:model_construction}
\end{figure}

Figure \ref{fig:model_construction} illustrates the Bayesian network structure and the particle filtering process. The left part of the figure shows the segmentation of the gesture area into different zones, each associated with a coil. The right part depicts the Bayesian network used for state transitions in particle filtering.

By integrating Bayesian networks into the particle filtering framework, our model achieves accurate gesture recognition and reliable trajectory tracking, enhancing the capabilities of multi - coil wireless charging pads for human - machine interaction.
\section{Experimental Verification}
\begin{figure}[h]
    \centering
    \includegraphics[width=0.4\textwidth]{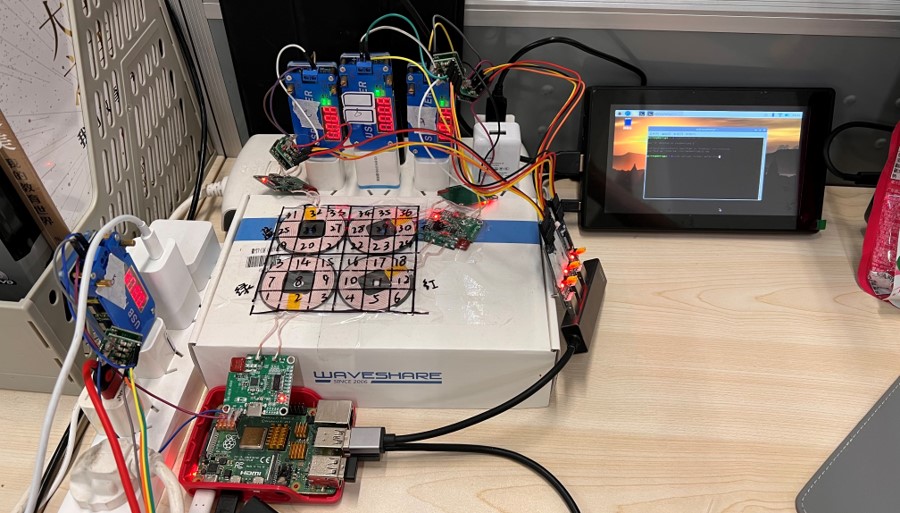} 
    \caption{Real - world experimental scenario}
    \label{fig:exp_scene}
\end{figure}
To validate the effectiveness of the proposed human - machine interaction solution based on multi - coil wireless charging pads, a series of experiments were carried out. The real - world experimental scenario is depicted in Fig. \ref{fig:exp_scene}.

The experimental setup consists of a multi - coil wireless charging pad, which serves as the core device for gesture recognition by detecting current and power changes among different coils. The pad is connected to a control unit, which is responsible for data acquisition and preliminary processing, such as noise reduction and time - series sorting of the coil current data. A high - pass filter is then applied to further eliminate background noise, and the processed data is sliced for subsequent analysis.

A Bayesian network is constructed to establish the relationships between variables related to gesture actions. Particle filtering is utilized to infer the motion trajectory of the user's gestures. Through repeated experiments and optimizations, the reasoning accuracy of the Bayesian network has reached 73\%. This indicates that the proposed model has a certain degree of reliability in gesture recognition.

During the experiments, various types of gestures were tested, including simple one - dimensional movements and complex multi - dimensional actions. The experimental results show that the proposed solution can effectively identify different gestures in a wide range of application scenarios. Moreover, compared with traditional gesture recognition methods, it has advantages such as lower cost and better adaptability in different environments.

In the experimental verification of the human - machine interaction solution based on multi - coil wireless charging pads, we obtained the following two key result figures.

\begin{figure}[h]
    \centering
    \subfloat[continuous distribution]{\includegraphics[width=0.24\textwidth]{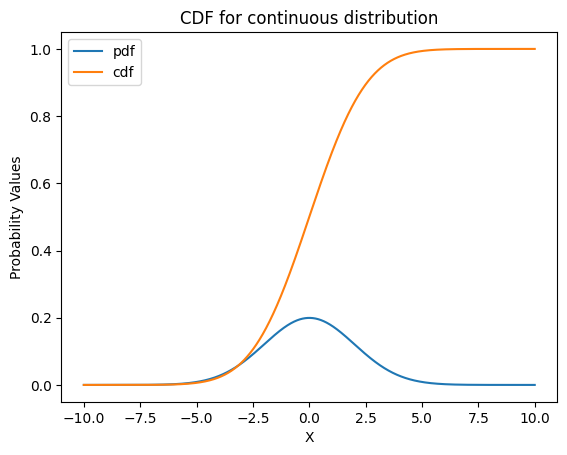}}
    \subfloat[Histogram Plot]{\includegraphics[width=0.24\textwidth]{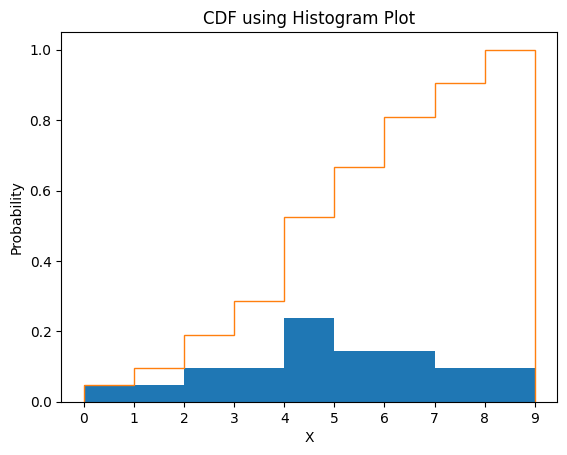}}
    \caption{Experimental result figures}
    \label{fig:exp_result_figures}
\end{figure}

The first sub - figure in Fig. \ref{fig:exp_result_figures} shows the probability density function (pdf, the blue curve) and the cumulative distribution function (cdf, the orange curve) for a continuous distribution. In the context of our experiment, the peak of the pdf curve in a specific region (such as near $x = 0$) may imply that in the related gesture recognition or data - processing process, the probability of occurrence of certain specific states or characteristic values is relatively high. This is crucial for understanding the distribution characteristics of gesture data and the central tendency in the recognition process. For example, when analyzing the coil current or power - change data corresponding to different gestures, if a certain range of values appears frequently, it may mean that the data within this range is closely related to common gestures.

The S - shaped trend of the cdf curve provides us with information about the cumulative probability of data. It indicates that as a certain variable (such as time, a certain quantitative index of gestures, etc.) increases, the cumulative probability of related events occurring gradually rises to approach 1. This helps us determine the probability of specific gestures or interaction behaviors occurring under different thresholds, providing a probabilistic basis for subsequent system decision - making and responses.
The second sub - figure in Fig. \ref{fig:exp_result_figures} intuitively presents the distribution and accumulation of experimental data through the combination of a histogram (blue bars) and a cumulative distribution curve (orange broken line). The differences in the heights of different intervals in the histogram reflect the frequency distribution of gesture data in each interval. For example, the relatively high data frequency in the interval $x = 4 - 5$ may mean that in the experiment, the gesture features corresponding to this interval (such as the gesture actions corresponding to a specific coil - current change range) are more common.
The cumulative distribution curve further shows the change in the cumulative probability of data with the $x$ - value. The increase in amplitude of each step corresponds to the frequency of data in the corresponding interval, enabling us to clearly understand the cumulative degree of data at different stages. 

\begin{figure}[htbp]
    \centering
    \subfloat[Bayesian Network Structure Learning]{\includegraphics[width=0.24\textwidth]{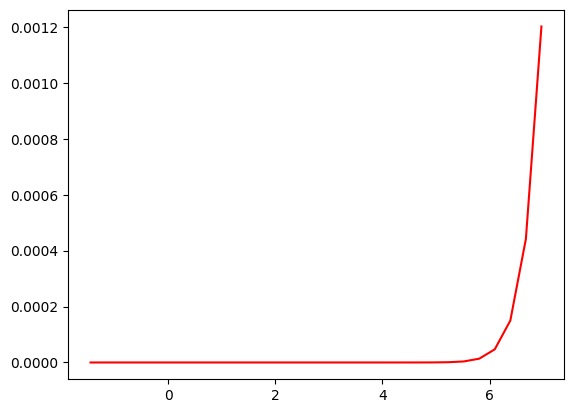}}
    \hfill
    \subfloat[Particle Filtering Parameter Ablation Test]{\includegraphics[width=0.23\textwidth]{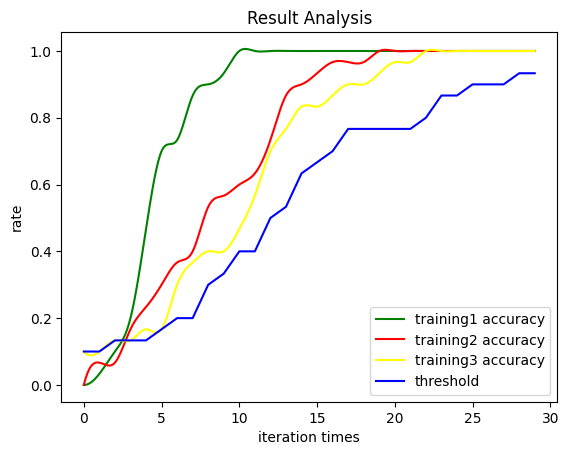}}
    \caption{Ablation Test Results}
    \label{fig:ablation_study}
\end{figure}

This study conducted ablation experiments to verify the effectiveness of each component in the human-machine interaction solution based on multi-coil wireless charging pads. Figure \ref{fig:ablation_study} presents two key ablation results. The first graph indicates that when using search-and-score methods (such as the K2 algorithm, etc.) for Bayesian network structure learning, the network structure scoring function rises significantly with the increase of the variable $x$, especially when $x > 6$, where the score increases sharply. This suggests that within this range of variables, the network structure significantly enhances the model's performance. The second graph shows the accuracy curves under different resampling thresholds, weight thresholds, and particle counts. It can be observed that different training sets exhibit different accuracy growth trends during the iteration process. As the number of iterations increases, the accuracy of each training set gradually improves and tends to stabilize. These results indicate that choosing appropriate resampling thresholds, weight thresholds, and particle counts is crucial for the performance of the particle filter. By adjusting these parameters, we can optimize the particle filter algorithm, thereby improving the model's accuracy and robustness.

Through the structure learning of the Bayesian network and the parameter ablation test of the particle filter, we verified the effectiveness of each component in the proposed human-machine interaction solution. The experimental results show that by optimizing the network structure and particle filter parameters, the performance of the model can be significantly enhanced. These findings not only validate the effectiveness of our method but also provide valuable insights for its future application in broader scenarios.



\begin{figure}[htbp]
    \centering
    \subfloat[Accuracy Comparison]{\includegraphics[width=0.24\textwidth]{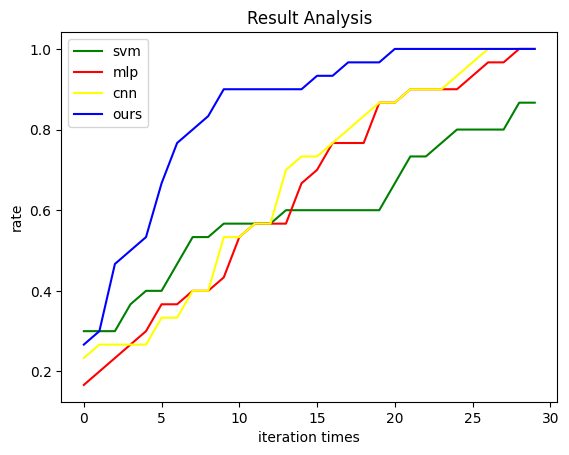}}
    \hfill
    \subfloat[Error Rate Comparison]{\includegraphics[width=0.24\textwidth]{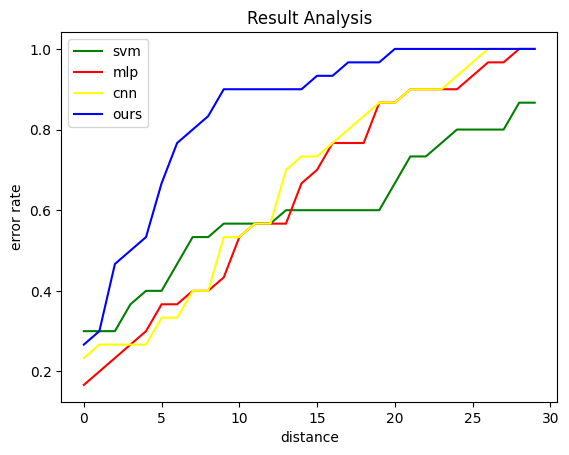}}
    \caption{Comparative Experiment Results}
    \label{fig:comparison}
\end{figure}

To validate the effectiveness of our proposed human-machine interaction solution based on multi-coil wireless charging pads, we conducted a series of comparative experiments. Figure \ref{fig:comparison} shows the performance comparison of our method with several commonly used machine learning models (SVM\cite{subha2024iacids3}, MLP\cite{gayathri2017iccsc}, CNN\cite{li2020dtcse}) in terms of accuracy and error rate.

As can be seen from Figure \ref{fig:comparison}, our method (blue line) achieves a higher accuracy rate with fewer iterations and continues to rise as the number of iterations increases, ultimately surpassing other methods. This indicates that our method has a faster convergence speed and higher accuracy when processing human-machine interaction data.

In the error rate comparison, our method also performs well, with its error rate consistently lower than other comparative methods, meaning our method has smaller prediction errors and thus better generalization ability and robustness.

Combining the results of accuracy and error rate comparisons, our method demonstrates superior performance in the human-machine interaction solution, not only outperforming other methods in terms of accuracy but also showing better stability and generalization capability of the model. This result further proves the effectiveness and reliability of our method in practical applications.

\section{conclusion}
This paper presents a low-cost, high-precision human-machine interaction solution based on multi-coil wireless charging pads. By monitoring changes in coil currents and utilizing Bayesian networks and particle filtering for gesture recognition, the experimental results validate the effectiveness of the proposed solution.
Compared with SVM, MLP, and CNN models, our method demonstrates superior performance in terms of accuracy and error rate, showing faster convergence and higher precision. This indicates that our approach has better performance when processing human-machine interaction data.
In summary, the proposed solution has broad application potential in smart homes, office automation, and public spaces, providing new directions for the multifunctional application of wireless charging pads and the development of human-machine interaction technology. Future work will focus on optimizing the algorithm and exploring more application scenarios.
\bibliographystyle{IEEEtran}
\bibliography{ssb}
\end{document}